\documentclass[preprint,showpacs,preprintnumbers,amsmath,amssymb]{revtex4-1}
\usepackage{graphicx}
\usepackage{dcolumn}
\usepackage{bm}
\usepackage{caption}
\usepackage{subcaption}
\newcommand{\be}{\begin{equation}}
\newcommand{\ee}{\end{equation}}
\newcommand{\bse}{\begin{subequations}}
\newcommand{\ese}{\end{subequations}}
\newcommand{\bary}{\begin{eqnarray}}
\newcommand{\eary}{\end{eqnarray}}
\newcommand{\bpmat}{\begin{pmatrix}}
\newcommand{\epmat}{\end{pmatrix}}
\newcommand{\bwt}{\begin{widetext}}
\newcommand{\ewt}{\end{widetext}}
\newcommand{\pee}{P_{ee}}
\newcommand{\peeb}{P_{{\bar e}{\bar e}}}
\newcommand{\pem}{P_{e\mu}}
\newcommand{\pemb}{P_{\bar e\bar \mu}}
\newcommand{\pet}{P_{e\tau}}
\newcommand{\petb}{P_{{\bar e}{\bar \tau}}}
\newcommand{\pmt}{P_{\mu\tau}}
\newcommand{\pmtb}{P_{{\bar \mu}{\bar \tau}}}

\begin{document}


\title{Oscillation of high energy neutrinos in  Choked GRBs}
\author{ Andr\'es Felipe Osorio Oliveros$^{1,2}$, Sarira Sahu$^{3}$, Juan
  Carlos Sanabria$^{1}$ 
}
\affiliation{
$^{1}$Universidad de Los  Andes, Bogota, Colombia\\
$^{2}$Fermi National Accelerator Laboratory, Batavia, IL 60510\\
$^{3}$Instituto de Ciencias Nucleares, Universidad Nacional Aut\'onoma de M\'exico, 
Circuito Exterior, C.U., A. Postal 70-543, 04510 Mexico DF, Mexico\\
}


\begin{abstract}

It is believed that choked gamma-ray bursts (CGRBs) are the potential
candidates for the production of high energy neutrinos in GeV-TeV energy
range. These CGRBs out
number the successful GRBs by many orders. So it is important to observe neutrinos
from these cosmological objects with the presently operating neutrino
telescope IceCube.  We study the three flavor neutrino oscillation of
these high energy neutrinos in  the presupernova star environment which is responsible for the 
CGRB. For the presupernova star we consider three different 
models and calculate
the neutrino oscillation probabilities, as well as neutrino flux on the
surface of these star. The matter effect modifies the neutrino flux of
different flavors on the surface of the star.  We have also calculated
the flux of these high energy neutrinos on the surface of the
Earth. We found that for neutrino energies below $\le 10$ TeV the
flux ratio does not amount to 1:1:1, whereas for higher energy
neutrinos it does.

\end{abstract}

\pacs{14.60.Pq; 98.70.Rz}
\keywords{Neutrino oscillation; Choked GRB}
\maketitle

\section{Introduction}
\label{intro}

The long Type ($\ge 2$ s) of gamma-ray bursts (GRBs) constitute about 3/4 of the total
observed GRBs. The observed correlations of the following GRBs with
supernovae (SNe), 
GRB980425/SN 1998bw \cite{Galama:1998ea},GRB 030329/SN 2003dh \cite{Mazzali:2003np}, 
GRB 031203/SN 2003lw\cite{Thomsen:2004sp} and GRB 021211/SN 2002lt\cite{DellaValle:2003uj}
show that core collapse of massive stars are related to long type of GRBs.
These mostly occur in star forming regions\cite{Woosley:2006fn,Zhang:2009uf}.  
The core collapse of massive stars 
resulting in a relativistic jet which breaks through the stellar envelope is 
a widely discussed scenario for gamma-ray burst production. In this scenario,
the gamma-rays are produced by synchrotron and/or inverse-Compton scattering of
Fermi accelerated electrons in optically thin shock, when the jet has emerged 
out of the stellar envelope. These same shocks are also responsible
for the acceleration of 
the protons into relativistic velocities and collisions of these
with the MeV photons  produce neutrinos of energy $\sim 100$
TeV\cite{Waxman:1997ti}.

The formation of jets may be a common phenomena in a collapsar scenario
and depending on the composition (baryon load) 
all the jets may not have sufficient energy and momentum to punch trough the
stellar envelope. From the observed rate of GRBs, only about $\le 10^{-3}$ core
collapse supernovae produce highly relativistic jets (Lorentz factor
$\Gamma \ge 100$)  which can penetrate through the
envelope and produce GRB events in an optically thin environment outside the star\cite{Berger:2003kg}.
On the other hand,  a large fraction of them will fail to emerge out of the envelope, which may
give rise to orphan radio afterglow instead of $\gamma$-ray emission.
However when the jets are making their way through the star
they can accelerate protons to energy $\ge 10^5\,{\rm  GeV}$ through internal shocks well inside 
the stellar envelope. Also the buried jet produces thermal X-ray at
$\sim 1\, {\rm keV}$ which acts
as the target for the delta-resonance to produce $\ge$ 5 TeV
energy neutrinos through photopion production, which  penetrates
through the envelope\cite{Meszaros:2001ms}. There can also be neutrino production due to $pp$ and $pn$ collisions
involving relativistic protons from the buried jet and the thermal nucleons from the jet and the
surrounding\cite{Razzaque:2004yv}. However, 
when the jets are still making their way through the star, precursor TeV neutrinos are inherent
to the collapse of a massive star irrespective of whether GRB is produced or not. 
Forward moving jets which are unable to emerge out of the envelope may be rich in baryons
which make them mildly relativistic, but at the same time efficient producers of TeV neutrinos 
through photomeson interaction. The detection of low luminosity GRB 060218\cite{Campana:2006qe} suggests that 
their number (low luminosity GRBs and dark GRBs) may be 
quite large compared to the high luminosity GRBs, which can probably contribute more to 
the TeV neutrino background\cite{Liang:2007} than the high luminosity ones and could be
detectable by present
day neutrino telescopes (e.g. in IceCube)  which can shed more light on the nature of the
central engine, as well as the acceleration mechanism of high energy
cosmic rays in the presupernova star\cite{Razzaque:2003uv,Razzaque:2005bh}.
From an individual collapse/GRB burst at a distance $z\sim 1$, about 0.1-10 upward
going muon events can be detected in a ${\rm km}^3$ detector\cite{Meszaros:2001ms}.

These high energy neutrinos propagating through the presupernova star
with a heavy envelop can oscillate to other flavors due to the matter
effect. In fact the mater effect is well known for neutrinos
propagating in the sun,
supernova as well as in the early Universe. 
In optically thick hidden sources where gamma-rays are not observed directly and 
$\sim {\rm TeV}$ neutrinos are produced due to $p\gamma$, $pp$ and $pn$ collision, the flux ratio at 
the production site and on the surface of the star may be different due to matter effect on their 
oscillation.  Recently for two neutrino flavors it is shown that for the choked GRBs,
the multi-TeV neutrino signals proposed by M\'esz\'aros and Waxman
\cite{Meszaros:2001ms} can undergo substantial resonance oscillation
before escaping from the He envelope if the neutrino oscillation
parameters are in the atmospheric neutrino oscillation range.
This would alter the neutrino flavor ratio escaping from the
stellar envelope, and subsequently the  detected flavor ratio 
on Earth\cite{Sahu:2010ap}. So in this context it is important to
study the matter effect of the presupernova star on the oscillation of
high energy neutrinos emerging out of it.

The neutrino oscillation in vacuum and matter has been discussed
extensively for solar, atmospheric as well as accelerator and reactor
experiments. Models of three flavor neutrino oscillations in constant matter
density\cite{Barger:1980tf,Kim:1986vg,Zaglauer:1988gz}, linearly
varying density\cite{Petcov:1987xd,Lehmann:2000ey} and exponentially 
varying density\cite{Osland:1999et} have been studied.
In Ref.\cite{Ohlsson:1999xb} 
T. Ohlsson and H. Snellman 
have developed an analytic formalism for the
oscillation of three flavor neutrinos in the matter background with varying
density, where they use the plane wave approximation for the
neutrinos (henceforth we refer to this as OS formalism).
Here  the evolution operator and the transition probabilities
are expressed as functions of the vacuum mass square differences,
vacuum mixing angles and the matter density parameter. As application
of the above formalism, the authors have studied the neutrino
oscillations traversing the Earth and the Sun for constant, step-function
and varying matter density
profiles\cite{Ohlsson:1999um,Ohlsson:2001et}. 
To handel the varying density,
the distance is divided into equidistance slices and in each slice the
matter density is assumed to be constant. In these calculations they have
considered the CP phase to be real by taking the phase factor
$\delta_{CP}=0$ so that the neutrino mixing matrix is real.

Although the OS formalism is simple and used for low energy neutrino oscillation, so far it has not
been used to study the propagation of high energy neutrinos neither in
the stellar envelope where the density is high nor in  the Earth. 
In the present work we are using this formalism to study the three
flavor high energy (energy in the range 100 GeV to 100 TeV )  neutrino oscillation
when traversing the presupernova star medium and reaching to the Earth
by undergoing vacuum oscillation in the intergalactic medium.

The paper is organized as follow: In Sec.\ref{sec:2}  we review the OS
formalism used for the calculation of neutrino oscillation
probabilities in a medium.In Sec. \ref{sec:3}   we discuss about the
presupernova star models to explain the choked GRBs. The discussion of
our results are given in Sec. \ref{sec:4}. followed by a summary in Sec. \ref{sec:5}.

\section{Neutrino oscillation Formalism}
\label{sec:2}

In this section we shall summarize the formalism used by OS\cite{Ohlsson:1999xb} 
for the calculation of the oscillation probability of the three active
flavors. A flavor neutrino state is a linear superposition of mass eigenstate and is
given as 
\be
 |\nu_{\alpha}\rangle = \sum^{3}_{a=1} U^*_{\alpha a}|\nu_{a}\rangle , 
\label{flavormatter}
\ee
where $\alpha = e, \mu, \tau$ (flavor eigenstates) and $a = 1,2,3$ (mass
eigenstates). The matrix $U_{\alpha a}$, is the three neutrino mixing
matrix given by,
\bwt
\be
  U = 
\bpmat
U_{e1} & U_{e2} & U_{e3}\\
U_{\mu 1} & U_{\mu 2} & U_{\mu 3}\\
U_{\tau 1} & U_{\tau 2} & U_{\tau 3}\\
\epmat
=
\bpmat
 c_{13} c_{12} & c_{13} s_{12} & s_{13} e^{-i\delta_{cp}} \\
 -s_{12} c_{23} - c_{12} s_{23} s_{13} e^{i\delta_{cp}} &  c_{12} c_{23} - s_{12} s_{23} s_{13} e^{i\delta_{cp}}  & s_{23} c_{13}\\
 s_{23} s_{12} - c_{23} s_{13} c_{12} e^{i\delta_{cp}} &  - s_{23} c_{12} - s_{13} s_{12} c_{23} e^{i\delta_{cp}} & c_{23}c_{13}\\
\epmat,
\label{pmns}
\ee
\ewt
where $c_{ij} \equiv \cos\theta_{ij}$ and $s_{ij} \equiv
\sin\theta_{ij}$ for $i,j = 1,2,3$. The neutrino mixing angles 
are $\theta_{12}, \theta_{13}$ and $\theta_{23}$. The $\delta_{CP}$ is
the CP violating phase. As the CP violation is not observed in the
neutrino sector, we put this phase  $\delta_{CP}=0$ in our
calculation.

While the neutrinos travel from the production point to the detection point, the flavor ratios 
will evolve as a result of their oscillations. These neutrinos will go through vacuum and matter during their 
propagation. In vacuum, the Hamiltonian that described the propagation of the neutrinos in the mass 
eigenstate basis  is described by 
\be
H_{m} = 
\bpmat
E_{1} & 0 & 0 \\
0 & E_{2} & 0 \\
0 & 0 & E_{3} \\
\epmat,
\label{Hm}
\ee
where $E_{i}$, for $i=1,2,3$ refer to the energy of each neutrino mass eigenstate. This Hamiltonian can be 
written in the flavor basis through the unitary transformation
described by the matrix $U$ from equation \eqref{pmns}, as
\be
 H_{f} = UH_{m}U^{-1}.
\ee
 In matter, the $\nu_e$ and ${\bar \nu}_e$
will interact through both charge and neutral current, whereas
$\nu_{\mu}$ and $\nu_{\tau}$ and their anti-neutrinos will interact
through neutral current only. So for the oscillation of electron
neutrino (or anti-neutrino) to other flavor, only the charge current
term will contribute. For the oscillation of $\nu_{\mu} \leftrightarrow
\nu_{\tau}$, there is no contribution from the matter up to leading order.
Thus the matter contribution to the Hamiltonian in the flavor basis
can be expressed as 
\be
V_{f} = 
\bpmat
A & 0 & 0 \\
0 & 0 & 0 \\
0 & 0 & 0 \\
\epmat,
\label{HV}
\ee
where A represents the potential due to the interaction of $\nu_{e}$/$\bar{\nu_{e}}$ with matter and is given by  
\be
 A = \pm \sqrt{2} G_{F} N_{e} =\pm\sqrt{2} G_{F} \frac{\rho}{m_{N}},
 \label{potential}
\ee
where $m_{N}$ is the nucleon mass, $G_{F}$ is the Fermi 
coupling constant and $\rho$ is the matter density.
The neutral current contribution to the potential is the same for all the
three neutrinos, so here we do not take that into account. 
In the mass basis, the total Hamiltonian is given by
\bary
 {\cal H}_{m} &=& H_{m} + U^{-1} V_{f} U\\ \nonumber
                    &=& H_{m}+V_{m}.
 \label{htot}
\eary
The total Hamiltonian in the flavor basis is written as
\be
{\cal H}_{f}=H_f + V_f.
\ee
For neutrino propagation in a medium, the Hamiltonian is not diagonal,
neither in the mass basis nor in the flavor basis, so one has to
calculate the evolution operator in any of these basis.

In the mass basis, the evolution of the state at a later time $t$ will be obtained by solving the
Schr\"{o}dringer  equation 
\be
 i\frac{d |\nu_{a} (t) \rangle}{dt} = {\cal H}_{m} |\nu_{a} (t)\rangle,
\ee
 and the solution to this equation can be expressed in terms of the evolution operator as
\bary
 |\nu_{a} (t)\rangle &=& e^ {-i {\cal H}_{m} t} |\nu_{a} (0)\rangle \\ \nonumber 
                    &=& U_{m}(t) |\nu_{a} (0)\rangle,
 \label{evol}
\eary
where $U_{m}(t)$ is the evolution operator in the mass basis and in
the flavor basis this can be written as 
\be
U_{f}(t) = U U_{m}(t) U^{-1}.
\ee
As neutrinos are relativistic, we can replace $t$ by the path length $L$,
where we use the natural units $c=1$ and $\hbar=1$ . To obtain $U_m
(L)$ we have to evaluate the exponential of the matrix $-i{\cal H}_m
L$ and also introduce a traceless matrix $T$ defined as
\be
 T = {\cal H}_{m} - \frac{tr({\cal H}_{m} ) I}{3}.
\label{matT}
\ee
The trace of the Hamiltonian in the mass basis is
\be
tr\left ( {\cal H}_m\right )=\sum^3_i E_i+ A.
\ee
For the square matrix, $T$, its exponential is given by
\be
e^{T}=\sum_{n=0}^{\infty} \frac{T^n}{n !}.
\ee
Using Cayley-Hamilton's theorem, this infinite sum can be expressed as
a finite sum and can be given by
\be
e^{T}=\sum_{n=0}^{N-1} a_n {T^n},
\ee
where $N$ is the dimension of the matrix $T$.  Here we have $N=3$ and
the matrix $T$ has three  eigenvalues $\lambda_i$ with $i=1,2,3$. The
solution of the characteristic equation of the $T$ matrix will be
obtained by solving
\be
\lambda^3+c_2\lambda^2+c_1\lambda+c_0=0.
\ee
The coefficients of $\lambda$ are given as
\be
c_2=-tr(T)=0, \,\,c_0=-{\rm det}(T),
\ee
and
\be
c_1=T_{11}T_{22}-T^2_{12}+T_{11}T_{33}-T^2_{13}+T_{22}T_{33}-T^2_{23}.
\ee
The eigenvalues are given by
\bary
\lambda_1&=& \frac{X}{2^{1/3} 3^{2/3}} -\frac{\left (\frac{2}{3}\right
)^{1/3}c_1}{X},\nonumber\\
\lambda_{2,3}&=& \frac{(1\pm i\sqrt{3}) c_1}{2^{2/3} 3^{1/3} X}
-\frac{(1\mp i\sqrt{3}) X}{2\times 2^{1/3} 3^{2/3}},
\eary
with
\be
X=\left (  
\sqrt{3} \sqrt{ 4 c_1^3 + 27 c_0^2}-9c_0 
\right )^{1/3}.
\ee
With the use of the above, the evolution operator in the mass basis can
be written as
\bary
U_{m}(L) &=& e^{-i {\cal H}_{m} L} \\ \nonumber
         &=& \phi \sum^{3}_{a=1} e^{-iL\lambda_{a}} \frac{\left[ (\lambda^{2}_{a} + c_{1}) I + \lambda_{a} T + T^{2} \right]}{3
           \lambda^{2}_{a} 
+ c_{1}},
\label{uevol2}
\eary
where
$\phi=e^{-iL\, tr({\cal H}_m) I/3}$ is a
complex phase
factor and $I$ is the identity matrix. The evolution operator in the flavor basis 
is given by 
\bary
U_{f} (L) &=& e^{-i {\cal H}_{f} L} \\ \nonumber
         &=& U e^{-i {\cal H}_{m} L} U^{-1} \\ \nonumber
         &=& \phi \sum^{3}_{a=1} e^{-iL\lambda_{a}} \frac{\left[ (\lambda^{2}_{a} + c_{1}) I + \lambda_{a} \tilde{T} + \tilde{T}^{2} \right] }{3 \lambda^{2}_{a} + c_{1}},
 \label{evolflavor}
\eary
where $\tilde{T} = UTU^{-1}$ and  $\tilde{T^2} = UT^2U^{-1}$. 
The probability 
of flavor change from $\alpha$ to $\beta$ due to neutrino oscillation through a distance L can be given by
\bary
&&P_{\nu_{\alpha}\rightarrow{\nu_{\beta}}}(L) \equiv 
 P_{\alpha \beta}(L) = |\langle \nu_{\beta} | U_{f}(L) | \nu_{\alpha}
 \rangle|^{2} \nonumber\\
&=& \delta_{\alpha \beta} - 4 \underset { a < b}{\sum^3_{a=1} \sum^3_{b=1}}
P_{a}(L)_{\beta\alpha} P_{b}(L)_{\beta\alpha} \sin^2 x_{ab}, 
 \label{prob}
\eary
where we have defined
\be
P_{a}(L)_{\beta\alpha} =\frac{
(\lambda^2_a+c_1)\delta_{\beta\alpha} +
  \lambda_a {\tilde T}_{\beta\alpha} + {\tilde T^2}_{\beta\alpha}
}{3 \lambda^2_a + c_1}.
\ee
The matrices ${\tilde T}_{\beta\alpha}$ and ${\tilde
  T^2}_{\beta\alpha}$ are symmetric and defined as
\be
{\tilde T}_{\alpha\beta}={\tilde T}_{\beta\alpha}=\sum^3_{a=1} \sum^3_{b=1} U_{\alpha a}
U_{\beta b} T_{ab},
\ee 
and
\be
{\tilde T^2}_{\alpha\beta}={\tilde T^2}_{\beta\alpha}=\sum^3_{a=1} \sum^3_{b=1} U_{\alpha a}
U_{\beta b} T^2_{ab}.
\ee
Also we have defined the quantity
\be
x_{ab}=\frac{ (\lambda_a-\lambda_b) L}{2}.
\ee
The matrix T is written explicitly as 
\be
T_{ab}= 
\bpmat
T_{11}
&
A U_{e1} U_{e2} & A U_{e1} U_{e3} \\
A U_{e1} U_{e2} & 
T_{22} & A U_{e2} U_{e3}
\\
A U_{e1} U_{e3} & A U_{e2} U_{e3}  & 
T_{33}
\\
\epmat,
\label{tmatrix}
\ee
where the diagonal elements of the above matrix are given by
\be
T_{aa}=A U^2_{ea} + \frac{1}{3} (\sum^3_{b\neq a=1}E_{ab}-A).
\ee
Here $E_{ab}=-E_{ba}=E_a - E_b$ and
the energies satisfy the relation 
\be
E_{12}+E_{23}+E_{31}=0.
\ee
The neutrino oscillation probabilities satisfy the condition
\be
\sum_{\beta} P_{\alpha\beta}=1,\,\,\, {\rm for}\,\,\,
\alpha,\beta=e,\mu,\tau, 
\label{probsum}
\ee
and a similar condition is satisfied for anti-neutrinos which we
define as $P_{{\bar\alpha}{\bar\beta}}$.

Using the Eqs.(\ref{prob}) and (\ref{probsum}) we can calculate the probability of
transition from one flavor to another.  By substituting  $A=0$ in
Eq.(\ref{prob}) we get the vacuum probability. For matter with varying
density the distance $L$ can be discretized into small segments with a
constant density in each segment and use this formulation repeatedly
in each segment. By doing so we can study numerically the neutrino
oscillation in any type of density profile. For neutrinos traversing a
series of matter densities $\rho_i$ for $i=$ 1 to $n$, with their
corresponding thickness $L_i$, the total evolution operator is the
ordered product and is given as
\be
U_f(L)=\prod^n_i U_f(L_i),
\label{produf}
\ee 
where $\sum^n_i L_i=L$.
For rapidly changing
profiles one has to consider smaller segments to make sure that density is
almost constant in each of the segments.
In a series of papers by OS, this method has been applied for
different density profiles of the Sun and the Earth, to study the MeV energy
neutrino oscillation.

\section{Presupernova star models}
\label{sec:3}

The popular models for the long duration gamma-ray bursts (LGRBs) are
the core collapse supernova models called the {\it collapsar  models}, where the
core of a massive star collapses to form a black hole and drive an
ultra relativistic jet, which breaks out of the star\cite{MacFadyen:1999mk}. Due to the
rotation of the star, the mass distribution along the rotation axis is lower than
the equatorial region which helps to launch the relativistic jet along the rotation axis of the 
progenitor forming a black hole at the center. 
The jet is probably powered by the annihilation of
neutrino anti-neutrino pairs or some other electromagnetic process.
Even though the formation of jets may be a common phenomena in a collapsar scenario,
the mass density of the progenitor is quite high, so it is not so trivial that
all the formed outflow can always be collimated and punch through the stellar envelope as a jet.
Depending on the initial mass and metallicity, the presupernova star
can have different compositions and different radii which can not be
probed observationally. So for this reason we resort to models that
assume some standard processes considered by different authors,
and which seem to fit the observational data well.  A general model for the density of the progenitor 
of the SN is based on a power law, $\rho \propto r^{-n}$\cite{Matzner:1998mg}
and for a star this parametrization is valid 
only in a range of $r$, but should be a fairly good approximation for our calculation. 
The power index $n$ depends on the stage in wich the progenitor is. There are two commonly used 
values: $n=3$ and $17/7$.  The value $n=3$ \cite{Shigeyama:1990,Arnett:1991} resulting from the study of the SN1987A and 17/7 for a simple 
blast wave model \cite{Chevalier:1989}.
Also depending on the composition of the presupernova star, the spectra will have some specific features that allow
a classification of the SNe. The presupernova could be a red super giant (RSG) or a 
blue super giant (BSG), both having hydrogen envelope but different energy transfer mechanisms 
(convective or radiative respectively).

\begin{figure}[t!] 
  \centering
  \resizebox*{0.5\textwidth}{0.3\textheight}{\includegraphics{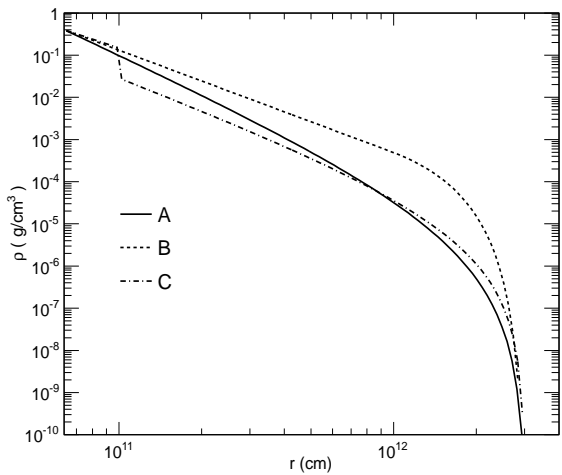}}
  \caption{Density profiles of the pressupernova star ( a blue
    supergiant) of models A, B and
    C with a radius $R_{*}=3\times 10^{12}$ cm. In these models, the
    high energy neutrinos are produced at a radius $r_j=10^{10.8}$ cm.}
  \label{presupernova_star_densityprofile}
\end{figure}


Based on above and with the fact that some of the  Long GRBs observed are related to SNe (Type
Ib/c), we are going to use three possible models as considered in \cite{Mena:2006eq}. 
The characteristics of these progenitor models are that  they are 
having an iron core of radius
$r_{Fe}\sim 10^9$ cm surrounded by a He core extending up to $r_{He}\sim 10^{11}$
cm where the density is $\rho_{He}\sim 10^{-3}\, g/cm^3$. In some cases a
hydrogen envelope surrounds the He core extending to $r_{H}\gtrsim
10^{13}$ cm with a density of
$\rho_{H}\sim 10^{-7}\, g/cm^3$ .
The presupernova stars which are believed to be the strong contender
for the Long GRBs are Type Ic SNe which have lost their hydrogen
envelopes as well as most of the He envelope before the explosion. These
objects are not interesting from the point of view of neutrino
oscillation because their radii are too small to have any appreciable
effect. On the other hand, the
presupernova stars with the He envelope (Type Ib SNe) and even the H
envelope (Type II SNe)  intact will be favorable for TeV neutrino
production as well as their oscillation in the stellar environment. We
consider three different presupernova models as shown in FIG. \ref{presupernova_star_densityprofile}.
In all these models the radius of the BSG is taken to be
$R_{*}\simeq 3\times 10^{12}$ cm. The jet evolves at 
a radius $r_j \simeq 10^{10.8}$ cm $< r_{He}$ and also the $\gtrsim$ 5 TeV neutrinos are produced at a point between
$r_j$ and $r_{He}$ so that the jet can acquire relativistic velocity on the surface of the He
envelope. The models are:
\begin{itemize}
 \item \textbf{Model A}
\be
\rho(r) =\rho_{\ast} \times \left(\frac{R_*}{r}-1\right)^3
\frac{\rm g}{\rm cm^{3}}.
\label{profileA}
\ee
This model represents a star with a radiative envelope. It has a polytropic structure 
with a polytropic index $n = 3$ and the characteristic density
$\rho_{\ast} = 4.0\times10^{-6}$ g/cm$^{3}$. 
This model is valid only 
in the region of the star that lies between the point were the jet is
produced and the star envelope i.e. $r_j\lesssim r\lesssim R_*$.
\item \textbf{Model B} 
\bary
\rho(r) =\rho_{\ast} \times
\left\{ \begin{array}{ll}
\left(\frac{R_{*}}{r}\right)^n & ;
10^{10.8}\,{\rm cm}
< r< r_b\\
\left(\frac{R_{*}}{r}\right)^n\frac{\left(r-R_{*}\right)^5}{\left(r_b-R_{*}\right)^5}
& ; 	r> r_b    
\end{array}     \right.    \frac{\rm g}{\rm cm^3} .\nonumber\\                
\label{profileB}       
\eary
This model is for a BSG with polytropic index $n$=17/7 and with
$r_b=10^{12}\, {\rm cm}$. The density 
has a power law behavior with a characteristic density $\rho_{\ast} = 3.4\times10^{-5}$ g/cm$^{3}$. This 
profile was obtained from the fit to SN1987A data.
\item \textbf{Model C}
\bary
\rho(r) = \rho_{\ast} {\cal A }\left(\frac{R_{*}}{r}-1\right)^{n_{eff}}
\frac{\rm g}{\rm cm^3}, 
\label{profileC}
\eary
where the parameters of model C are given as
\bary
\left(n_{eff},{\cal A}\right)=\left\{ \begin{array}{ll}
 (2.1,\, 20) & ; 10^{10.8}\, {\rm cm}<r <10^{11}\\
(2.5,\, 1.0) & ; r >10^{11}\, {\rm cm}.\\
\end{array} \right.  \nonumber\\
\eary
This model has two free parameters $n_{eff}$ and ${\cal A}$ that can be fitted to produce a drop in the density
after the helium core.  While the
parameter ${\cal A}$ can be used to set a density drop at the edge of the helium core, 
$n_{eff}$ gives the effective polytropic index \cite{Matzner:1998mg}. The characteristic
density here is
 $\rho_{\ast} = 6.3\times10^{-6}$ g/cm$^{3}$. The density profiles of
 all these three models are shown in FIG.
 \ref{presupernova_star_densityprofile}. 
\end{itemize}

\section{Results}
\label{sec:4}

We use the neutrino oscillation formalism of  OS given in Sec. II and calculate the neutrino oscillation
probabilities and fluxes in presupernova star models A, B and C
and also on the surface of the Earth after the neutrinos have
undergone vacuum oscillation in the intergalactic medium. For all these calculations we take the neutrino
energy in the range 100 GeV to 100 TeV
, although it may be difficult
to produce neutrinos with energy more than 10-20 TeV in the presupernova
star environment. 

We use the standard neutrino oscillation parameters obtained from
different experiments for analysis of our results. The neutrino  parameters
used are as follows:
\bary
&&\Delta m^2_{21}=8.0\times 10^{-5}\, eV^2, \,\,
\theta_{12}=33.8^{\circ}\,\,{\rm
  and}\,\,\theta_{23}=45^{\circ}\nonumber\\
&& \Delta m^2_{32}=3.2\times 10^{-3}\, eV^2, \,\,\,
\theta_{13}=8.8^{\circ}\,\,{\rm
  and}\,\,\delta_{CP}=0.
\eary
We also we take  $\theta_{13}=12^{\circ}$ 
to observe the variation in oscillation probability due to change in
this angle.


For the numerical
calculation, we divide the distance $(R_{*}-r_j)$ into small slices,
each with a constant density and calculate $U_f(L)$ for each
individual slices as discussed in
Eq.(\ref{produf}).  Afterward we use Eq.(\ref{prob}) to calculate the
probabilities. We let the high energy neutrinos propagate from the production point at
$r_j=10^{10.8}$ cm towards the surface of the star $R_*$, where we
calculate their survival and transition probabilities
$P_{\alpha\beta}$ and $P_{{\bar\alpha}\bar\beta}$, as well as fluxes. We consider two
different sets of parameters: $R_*=3\times 10^{12}\,{\rm cm}$
and $\theta_{13}=8^{\circ}$ (Set-I); $R_*=2.7\times
10^{12}\,{\rm cm}$ and $\theta_{13}=12^{\circ}$  (Set-II); to observe the variation in the probabilities at
different depth from the star surface and different $\theta_{13}$.

\begin{figure}[t!] 
{\centering
\resizebox*{0.5\textwidth}{0.3\textheight}
{\includegraphics{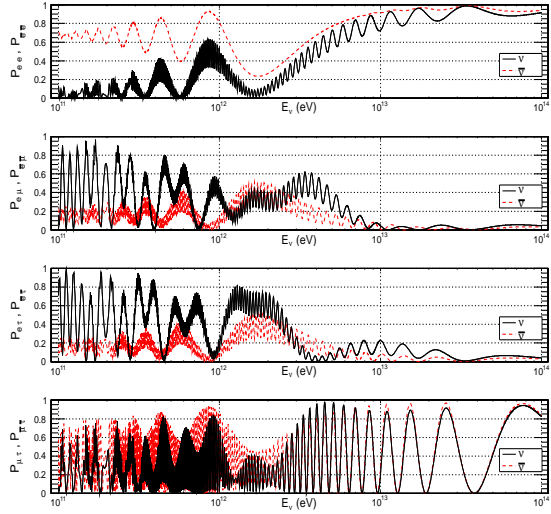}}
\par}
\caption{ The oscillation probabilities for neutrinos (black solid line) and
  anti-neutrinos (red dotted line) are plotted as functions of neutrino/anti-neutrino
energy $E_{\nu}$ for model A, Set-I where  first figure is for $\pee$ and
$\peeb$,  second figure is for $\pem$ and $\pemb$, third figure is for
$\pet$, $\petb$ and the fourth one is for $\pmt$ and $\pmtb$ respectively.
}
\label{presnmodelAI}
\end{figure}

\begin{figure}[t!] 
\vspace{0.3cm}
{\centering
\resizebox*{0.5\textwidth}{0.3\textheight}
{\includegraphics{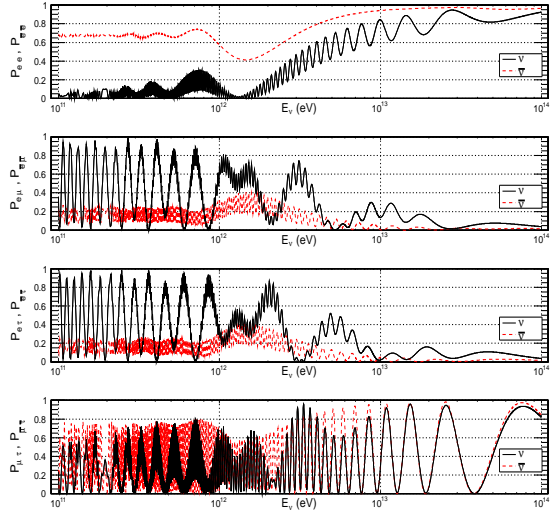}}
\par
}
\caption{This is same as FIG.\ref{presnmodelAI}  but for model-B.
(See the electronic edition of the Journal for a color version of
the figures and the color specification is same as FIG.\ref{presnmodelAI}).
} 
\label{presnmodelBI}
\end{figure}

\begin{figure}[t!] 
{\centering
\resizebox*{0.5\textwidth}{0.3\textheight}
{\includegraphics{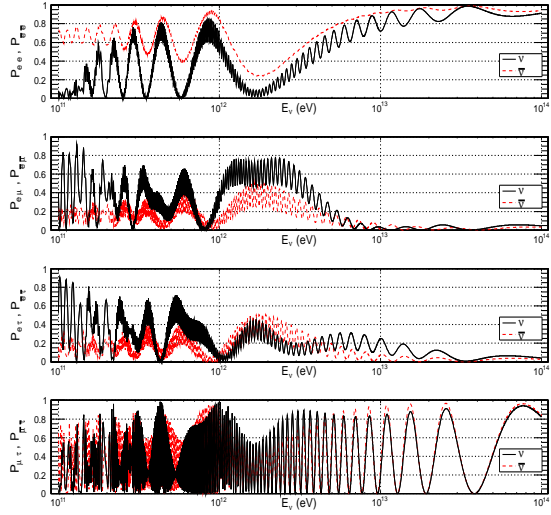}}
\par
}
\caption{This is same as FIG.\ref{presnmodelAI}  but for model-C.
}
\label{presnmodelCI}
\end{figure}

\begin{figure}[!t]
{\centering
\resizebox*{0.5\textwidth}{0.3\textheight}
{\includegraphics{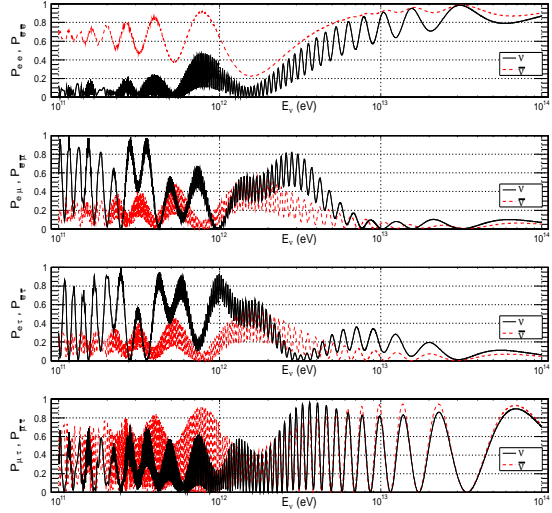}}
}
\caption{This is same as FIG.\ref{presnmodelAI} but for the parameter set-II.
} 
\label{presnmodelAII}
\end{figure}

\begin{figure}[!t]
{\centering
\resizebox*{0.5\textwidth}{0.3\textheight}
{\includegraphics{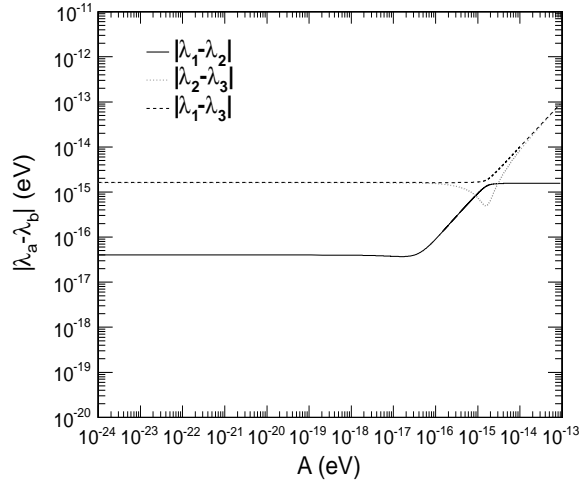}}
}
\caption{The energy difference is plotted as a function of potential
  energy A to look for the existence of resonance.
}
  %
\label{resoA}
\end{figure}

\begin{figure}[!t]
{\centering
\resizebox*{0.5\textwidth}{0.3\textheight}
{\includegraphics{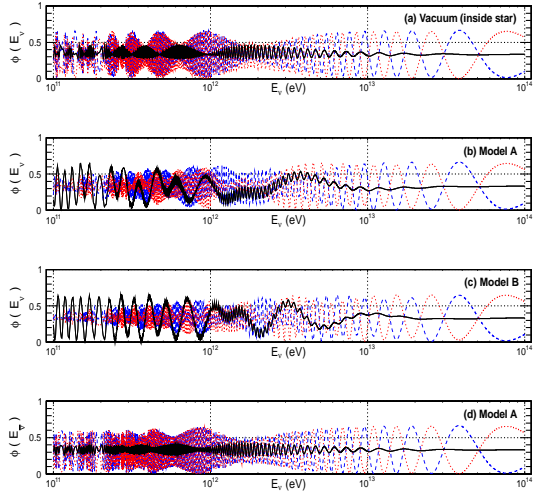}}
}
\caption{Neutrino and anti-neutrino fluxes on the surface of the
  star. In (a), (b) and (c) black solid line is for $\Phi_{\nu_e}$,
  blue dashed lines is for $\Phi_{\nu_\mu}$, and red dotted line is for
  $\Phi_{\nu_\tau}$. In (d) black solid curve, blue dashed curve
  and red dotted curve are for $\Phi_{{\bar\nu}_e}$, $\Phi_{{\bar\nu}_\mu}$
  and $\Phi_{{\bar\nu}_\tau}$ respectively.
}
\label{fluxPRESN}
\end{figure}

\begin{figure}[!t]
{\centering
\resizebox*{0.5\textwidth}{0.3\textheight}
{\includegraphics{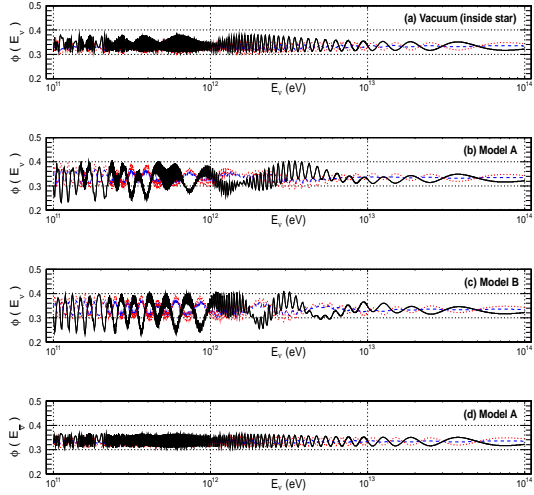}}
}
\caption{Same as FIG.\ref{fluxPRESN} but the flux is calculated on the surface of the
  Earth. 
}
\label{fluxEARTH}
\end{figure}

\begin{figure}[!t]
{\centering
\resizebox*{0.5\textwidth}{0.3\textheight}
{\includegraphics{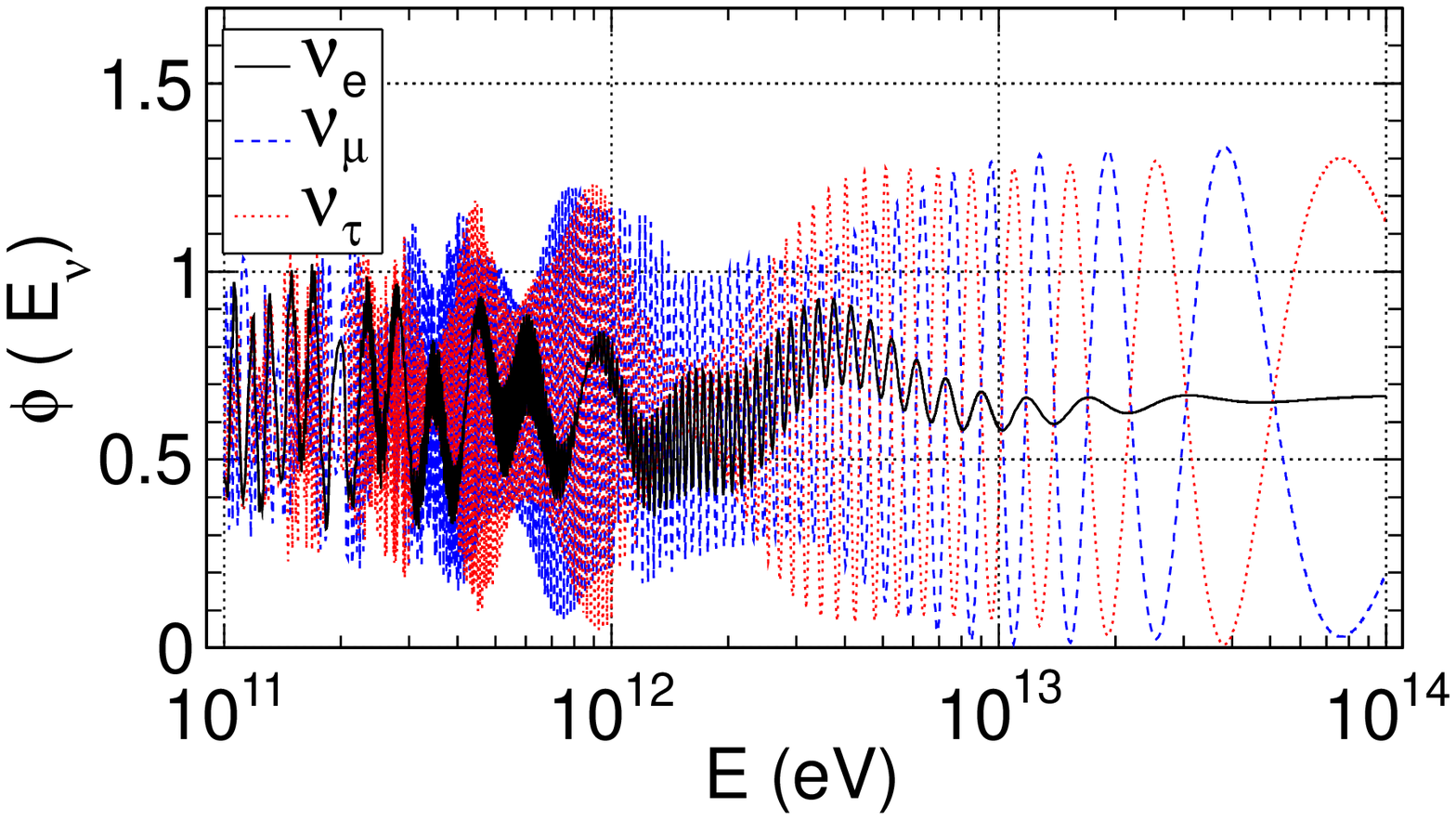}}
}
\caption{The total flux (neutrino+anti-neutrino) on the surface of the
  star, where the black solid, blue dashed and red dotted curves are for
  $\Phi_{\nu_e}$, $\Phi_{\nu_\mu}$ and $\Phi_{\nu_\tau}$ respectively.
}
  %
\label{totalfluxPRESN}
\end{figure}

\begin{figure}[!t]
{\centering
\resizebox*{0.5\textwidth}{0.3\textheight}
{\includegraphics{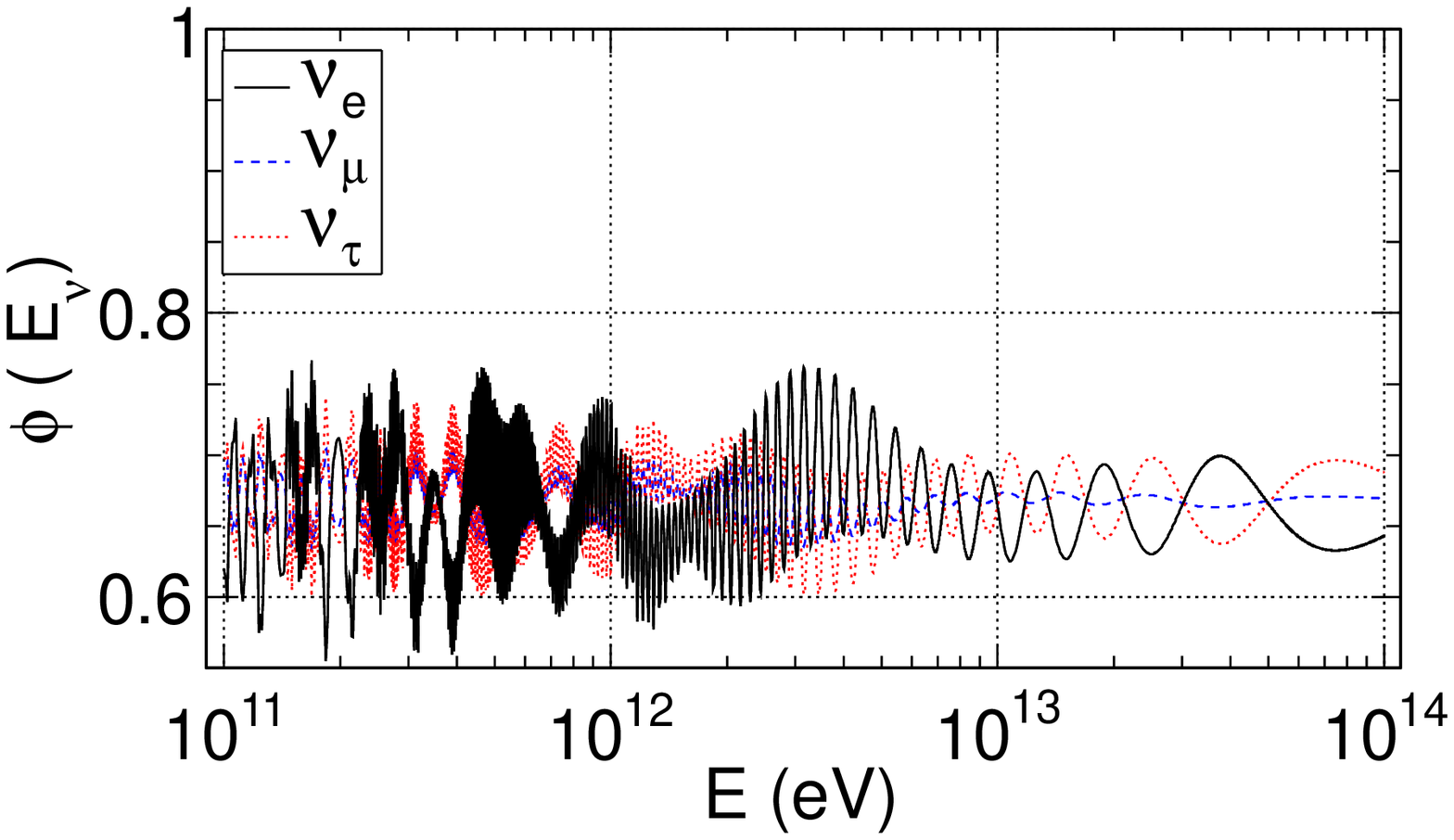}}
}
\caption{The total flux (neutrino+anti-neutrino) on the surface of the
  Earth, where the black solid, blue dashed and red dotted curves are for
  $\Phi_{\nu_e}$, $\Phi_{\nu_\mu}$ and $\Phi_{\nu_\tau}$ respectively.
}
\label{totalfluxEARTH}
\end{figure}

The survival and  transition probabilities of
$\nu$ and ${\bar\nu}$ for the parameter Set-I, for models A,
B and C are shown in FIGs. \ref{presnmodelAI}, \ref{presnmodelBI} and \ref{presnmodelCI}
respectively.  We have also plotted the variation in oscillation
probabilities for Set-II in FIG. \ref{presnmodelAII}. We observe that for a given
neutrino energy $E_{\nu}$, $\peeb$ is always above $\pee$ and the
probabilities are highly oscillatory for $E_{\nu}\lesssim 10^{12}$ eV. For 
$E_{\nu} \gtrsim 2\times 10^{12}$ eV, both $\pee$ and $\peeb$ increase
towards unity. While the increase in $\peeb$ is smooth, the
increase in $\pee$ is accompanied by a rapid oscillation, as shown in
these figures.  By reducing the $R_*$ to $2.7\times 10^{12}$ cm and $\theta_{13}=12^{\circ}$
i.e. Set-II (FIG. \ref{presnmodelAII} ), we observe that, although
there is variation of probability on the surface, the overall behavior
is exactly the same as the parameter Set-I shown in 
FIG.\ref{presnmodelAI}. So for our further discussion we only consider the parameter
Set-I for the analysis of our results. All the transition probabilities $\pem$, $\pet$, $\pmt$
$\pemb$, $\petb$ and $\pmtb$ are highly oscillatory in all the energy
ranges (mostly $\lesssim 10^{13}$ eV). For neutrino energy $E_{\nu} >
3\times 10^{12}$ eV  the transition probabilities  $\pem$, $\pet$, 
$\pemb$ and $\petb$ go to zero, which shows that the medium has almost no
effect on high energy 
neutrinos. On the other hand $\pmt$ and $\pmtb$ are highly oscillatory
in this energy range. The $\pem$ and $\pet$ are different from each
other due to the matter effect. Comparison of
$\pemb$ and $\petb$ shows that they oscillate with same amplitude, but
are out of phase by $180^{\circ}$.  The transition probabilities
$\pmt$ and $\pmtb$ are different for energies below $\sim 3\times
10^{12}$ eV, but are almost the same above this energy for all the
models A, B, and C (last  of FIGs. \ref{presnmodelAI},
\ref{presnmodelBI}, \ref{presnmodelCI} and \ref{presnmodelAII}) which
is due to the negligible medium effect on the high energy neutrinos.

The energy eigenvalues of the neutrinos in the matter are $\lambda_a$
and the energy difference $|\lambda_a-\lambda_b|$ is
related to the effective mass square difference 
\be
|\lambda_a-\lambda_b|=\frac{|{\tilde {\Delta  m}}^2_{ab}|}{2 E_{\nu}}.
\ee
In FIG.\ref{resoA}, for the illustrative purpose, we have taken
$E_{\nu}=1$ TeV with parameter Set-I and model-A to show the resonance
position as a function of matter potential A. It shows that
there is only one resonance position around $A\simeq 2\times 10^{-15}$
eV (between $ |\lambda_1-\lambda_2|$ and $ |\lambda_1-\lambda_3|$). 
By taking $E_{\nu}=10$ TeV we found that the resonance
position is almost the same. The anti-neutrinos will not satisfy the resonance condition
because of the change in the sign of the potential $A$ in $\lambda_a$.

In the mildly relativistic jet  internal shocks can develop and 
accelerate protons to very high energies. These protons
would interact with the $\sim$ keV thermal X-ray photons  
to produce TeV neutrinos via the process
$p+\gamma \rightarrow \Delta^+ \rightarrow n+ \pi^+
\rightarrow n+\mu^++\nu_{\mu}\rightarrow n+
e^++\nu_{\mu}+\nu_e+{\bar \nu_{\mu}}$. 
In the above process the standard neutrino flux ratio at the production point
is $\Phi^0_{\nu_e}:\Phi^0_{\nu_\mu}:\Phi^0_{\nu_\tau}=1:2:0$
($\Phi^0_{\nu_{\alpha}}$ corresponds to the sum of neutrino and
anti-neutrino flux at the source). 
The  flux observed at a distance is given by
\be
\Phi_{\nu_\alpha}=\sum_{\beta} \Phi^0_{\nu_\beta} P_{\alpha\beta}, \,\,
\alpha,\beta=e,\, \mu, \,\tau.
\ee

Using models A, B and C, we have calculated the normalized
fluxes of neutrinos and anti-neutrinos on the surface of the presupernova star as well as on the surface
of the Earth, which are shown in FIGs. \ref{fluxPRESN} and
\ref{fluxEARTH}. In these figures, (a) is the flux calculation in vacuum (where
the matter potential $A=0$ inside the star) and  the (b), (c) 
are for models A, B and (d) is for anti-neutrino fluxes for model A
respectively. The comparison of the vacuum oscillation i.e. (a)  with
the rest i.e. (b), (c) and (d), shows
that for $E_{\nu} \gtrsim 6\times 10^{12}$ eV both the fluxes are the
same, which signifies that above this energy the matter effect is
negligible and below this energy the normalized flux is oscillatory
and energy dependent.
Also the comparison of fluxes of neutrinos (b) and
anti-neutrinos (d) in model A below $E_{\nu} \gtrsim 6\times
10^{12}$ eV are different.

In FIGs. \ref{totalfluxPRESN} and \ref{totalfluxEARTH} we have shown
the total flux (neutrino and anti-neutrino) of $\nu_e$, $\nu_{\mu}$ and
$\nu_{\tau}$ on the surface of the star and on Earth. It can be
observed that the fluxes of different neutrinos are different on
the surface of the star. For $E_{\nu} \lesssim 4\times
10^{12}$ eV, the fluxes on Earth are also different but above this
energy the flux ratio is close to 1:1:1. 

\section{Summary}
\label{sec:5}

 It is observed that only a very small fraction of core collapse SNe 
are responsible for GRBs and majority of them fail to
produce $\gamma$-rays. This majority class are efficient
emitters of TeV neutrinos which can be observable in IceCube. Here we study the matter effect of the
presupernova star on the propagation of these
high energy neutrinos by taking into account all the three active
flavors and using the formalism developed by OS.
We observed that the neutrino oscillation
within the presupernova star depends on the neutrino energy and
alter their flux on the surface of the star.  We also calculated the
fluxes of these neutrinos on the Earth after they travelled the
intergalactic medium (vacuum oscillation) and found that low energy
neutrinos $E_{\nu} \lesssim 4$ TeV have different fluxes whereas above
this energy the flux ration is close to 1:1:1. So, possible detection of high
energy neutrinos with $E_{\nu}\lesssim 10$ TeV by IceCube can be
important due to the change in their fluxes when they propagate
through the choked environment of the presupernova stars. Also
detection of these neutrinos might shed more light on the type of progenitors and
the acceleration mechanisms of the high energy cosmic rays. 
As application and continuation of the present work, the
calculation of the neutrino induced  muon (track) to electron (shower)
ratio in IceCube is in preparation.

S.S. is thankful to Departamento de Fisica de Universidad de los
Andes, Bogota, Colombia, for their kind hospitality during his several
visits. We thank Karla Patricia Varela for helpful discussions.
This work is partially supported by DGAPA-UNAM (Mexico) Project 
No. IN103812.


\end{document}